\documentclass[twocolumn,aps,floatfix]{revtex4-2}

\usepackage[T1]{fontenc}
\usepackage[utf8]{inputenc}
\usepackage{microtype}
\usepackage{graphicx}
\usepackage{amsmath,amssymb}
\usepackage{hyperref}
\hypersetup{
  colorlinks=true,
  linkcolor=blue,
  citecolor=blue,
  urlcolor=cyan,
}
\usepackage{xspace}
\begin{document}
\title{Magnetically Guided Endothelial BioBots: A Next-Generation Strategy for Treating Complex Cerebral Aneurysms}

\author{Duong Le}
\affiliation{Department of Biomedical Engineering, University of Massachusetts Amherst, Amherst, MA, USA}
\email{duongle@umass.edu} 

\date{September 22, 2025}

\begin{abstract}
Cerebral aneurysms affect 3–5\% of the population, and rupture remains a major cause of stroke-related death and disability. Current treatment strategies, including surgical clipping, endovascular coiling, and flow diversion, have improved outcomes but still face important limitations. These shortcomings are especially evident in complex aneurysms, where irregular geometry and impaired biological healing make durable and safe repair particularly difficult. Clipping is invasive and often unsuitable for deep or posterior circulation lesions. Coiling is prone to recurrence due to compaction or incomplete occlusion, particularly in wide-neck or fusiform aneurysms. Flow diverters improve durability but rely on rigid metallic scaffolds that may fail to conform in tortuous vessels, compromise branch arteries, delay endothelialization, and require long-term dual antiplatelet therapy. This review and technical report introduces BioBots, a conceptual next-generation therapeutic platform designed to address these gaps. BioBots are biodegradable hydrogel carriers embedded with magnetic nanoparticles and coated with shear-primed endothelial progenitor cells. Delivered through microcatheters, they can be guided by external electromagnetic fields to assemble across aneurysm necks. Once localized, they create a conformal, geometry-adaptive endothelial patch that provides immediate antithrombotic and anti-inflammatory activity. As the hydrogel scaffolds degrade, only a stable endothelial lining remains, avoiding the long-term complications of permanent metallic implants. By combining microrobotic navigation with regenerative vascular biology, BioBots aim to accelerate occlusion, reduce recurrence, and promote durable vessel wall healing. This approach is particularly suited for complex aneurysms, where current therapies often fail to provide consistent or lasting outcomes.

\end{abstract}

\maketitle

\section*{Introduction}

Cerebral aneurysms are dilations of intracranial arterial walls that affect approximately 3--5\% of the general population \citep{ref1,ref2}. Although many remain unruptured, rupture can have devastating consequences: about 10\% of individuals with subarachnoid hemorrhage (SAH) caused by an aneurysm die before reaching medical attention; 25\% die within 24 hours, and 40--49\% die within three months \citep{ref1,ref3}. Survivors often sustain permanent neurological deficits \citep{ref1,ref3}. Risk of rupture depends on aneurysm size, location, patient sex, hypertension, and other factors, with aneurysms $\geq$ 10 mm carrying notably higher annual risk compared to smaller lesions \citep{ref3,ref4}. There is no universally accepted prevalence for `complex' aneurysms because complexity criteria differ across surgical and endovascular paradigms. Nevertheless, complex morphologies comprise a clinically significant subset: giant aneurysms represent $\sim$5\% of all intracranial aneurysms, fusiform and circumferential aneurysms account for $\sim$3--13\%, and the proportion categorized as wide-neck varies widely with the definition applied. Many bifurcation aneurysms, particularly at the MCA bifurcation, pose additional technical challenges \citep{ref4,ref5}.

Over the past several decades, treatment of cerebral aneurysms has evolved from open surgical clipping toward less invasive endovascular approaches, including coil embolization, stent-assisted coiling (SAC), and flow-diverting stents (flow diverters, FD) \citep{ref6}. Clipping remains a definitive option in many cases but carries higher morbidity in deep or posterior circulation aneurysms and is surgically demanding. Endovascular coiling is widely used, especially for saccular aneurysms, but it is associated with risks of coil compaction, residual aneurysm, and recurrence, particularly in wide-neck, large, or fusiform aneurysms \citep{ref7,ref8}. Flow diverters were developed to address these challenges by redirecting blood flow away from the aneurysm sac, promoting thrombosis, and encouraging neointimal growth across the aneurysm neck \citep{ref5}. However, they too have limitations: occlusion is often delayed, wall apposition can be incomplete in tortuous vessels, branch coverage may lead to ischemic complications, and long-term dual antiplatelet therapy (DAPT) is required to reduce thrombosis risk \citep{ref9,ref10}.

Despite these advances, many aneurysms, especially those with complex geometry, remain resistant to durable treatment. Recurrence rates after coiling or SAC are often in the 20--30\% range for certain morphologies \citep{ref7,ref11}. Flow diversion improves durability for many, but outcomes remain variable, and complications such as in-stent stenosis or branch vessel compromise are nontrivial \citep{ref10,ref12}. More importantly, none of the current therapies directly restore vascular biology: incomplete endothelialization, persistent inflammation, and extracellular matrix degradation often leave the parent artery structurally compromised even after apparent mechanical exclusion of the aneurysm \citep{ref33,ref34,ref35,ref36,ref37}. Thus, there remains a critical clinical gap: therapies that are both geometry-adaptive (able to conform to irregular or wide-neck lesions) and biologically active (able to re-establish endothelial function and promote vessel healing) are still lacking.

Encouraging precedents have emerged from related fields. Magnetizable flow diverters have demonstrated the ability to capture and retain endothelial cells \textit{in vivo}, accelerating endothelialization in animal models \citep{ref13}. Clinical-scale electromagnetic navigation systems capable of generating programmable magnetic fields across the human workspace have already been developed \citep{ref14}. Swarming nanorobots have been guided to vascular thrombi under flow, where coordinated mechanical activity enhanced fibrinolysis \citep{ref15}. Together these advances demonstrate the feasibility of magnetic endothelial targeting, collective microrobotic navigation, and eventual clinical translation.

Building on these foundations, we propose BioBots (magnetically guided, primed endothelial-cell-coated microrobotic systems) as a next-generation therapeutic platform for aneurysm repair. By combining magnetic guidance, biodegradable hydrogel scaffolds, and shear-primed endothelial biology, BioBots address the major limitations of existing devices: they are designed to adapt to irregular geometry, provide surfaces for endothelial adhesion even in areas lacking native substrate, and actively restore vascular integrity. While such a strategy could strengthen outcomes across aneurysm treatment broadly, it may be especially transformative in complex aneurysms, where current methods most often fail.

As this work represents a review and technical report, our goal is to synthesize evidence on the limitations of existing aneurysm therapies and recent advances in magnetic and bioengineered medicine, while introducing BioBots as a conceptual framework for overcoming the persistent challenges of complex aneurysm management.

\section*{Current Limitations and Reasons for Treatment Failure}

Surgical clipping excludes the aneurysm immediately by applying a metallic clip across its neck. This technique has long been considered the most definitive, with lower recurrence rates than endovascular approaches \cite{ref16}. However, clipping is invasive and technically demanding, requiring craniotomy and direct dissection that increase perioperative morbidity and extend recovery \cite{ref17}. Access to aneurysms in the posterior circulation or deep operative fields carries particularly high surgical risk \cite{ref18}. The success of clipping is closely linked to surgical expertise, and population-based studies show that older patients experience significantly higher complication rates and worse functional outcomes than younger patients \cite{ref19}. Therefore, although clipping remains effective for accessible aneurysms, concerns about complications limit its use in many patients. 

Endovascular coiling was developed as a less invasive alternative and is now widely adopted worldwide. The method provides short-term safety advantages but suffers from limited durability. Long-term angiographic series demonstrate recurrence rates of 20--30\%, with retreatment required in up to one-fifth of patients \cite{ref20,ref21}. Recurrence most often results from coil compaction under pulsatile flow, which reopens the aneurysm sac \cite{ref22}. This problem is particularly severe in wide-neck, fusiform, and giant aneurysms where stable coil packing is difficult to achieve. Even large clinical experiences confirm that many patients require repeat procedures during long-term follow-up \cite{ref23}. Balloon-assisted and stent-assisted coiling can improve packing density and short-term occlusion, but they increase procedure complexity and are associated with higher rates of ischemic complications \cite{ref24}. Evidence from the International Subarachnoid Aneurysm Trial demonstrated that while coiling offered lower disability and mortality in the first year compared with clipping, it was associated with higher rates of recurrence and retreatment at 10 to 18 years \cite{ref25,ref26}. These findings highlight the central weakness of coiling: it provides an initial reduction in risk but fails to consistently achieve long-term stability.

Flow diversion has been the most significant endovascular advance in the past two decades. These devices consist of braided stents with high metal coverage that redirect blood flow away from the aneurysm sac, leading to progressive thrombosis and gradual endothelialization across the device. Large pooled analyses report complete occlusion in 70--80\% of aneurysms at one year and more than 90\% by five years \cite{ref27,ref28}. However, occlusion is not immediate, and patients remain exposed to rupture risk for months while healing occurs \cite{ref29}. Technical deployment presents further problems. In tortuous or bifurcated vessels, achieving full wall apposition is often difficult, and up to one-quarter of cases require adjunctive balloon angioplasty to ensure proper device placement \cite{ref30}. Poor apposition leaves gaps between the device and the vessel wall, which can permit persistent aneurysm filling and increase the risk of retreatment \cite{ref31}. At the same time, coverage of branch vessels introduces another risk, as studies have documented ischemic complications including reduced flow or even occlusion in covered branches \cite{ref32}. Together, these challenges highlight how anatomical complexity and branch involvement limit the durability and safety of flow diversion.

The biological response to flow diversion also remains inconsistent. Although these devices are designed to promote uniform neointimal growth across the aneurysm neck, both imaging and histopathologic studies demonstrate that endothelial coverage can remain incomplete months after implantation \citep{ref33,ref34}. Patchy endothelialization leaves segments of device struts exposed to circulating blood, generating thrombogenic surfaces and disturbed flow that sustain aneurysm perfusion and instability, a mechanism linked to delayed rupture \citep{ref35}. Clinically, these biological shortcomings translate into necessary retreatment needs: while relatively small and simple aneurysms often achieve durable occlusion, approximately 5--10\% of cases still require additional intervention, and retreatment rates rise to 15--20\% in complex morphologies such as large, fusiform, or bifurcated aneurysms \citep{ref34}. Pathological analyses further demonstrate that aneurysm walls in these cases frequently show chronic inflammation, endothelial dysfunction, and extracellular matrix degradation \citep{ref36,ref37}, biological conditions that impair healing and compound the risk of rupture. Thus, because neither clipping, coiling, nor flow diversion addresses these biological dysfunctions, the parent artery remains structurally compromised even after treatment.

Moreover, in an effort to improve outcomes in complex cases, some centers have employed multiple overlapping stents to increase metal coverage. While this strategy can reduce inflow in theory, in practice it has not provided reliable occlusion of large or fusiform aneurysms. Studies show that multiple stents increase the risk of in-stent thrombosis, and do not significantly improve long-term occlusion compared with single devices \cite{ref38,ref39}. This approach highlights the broader limitation of mechanical strategies: adding more devices does not resolve the challenges posed by irregular geometry and diseased vessel biology.

In addition to these technical barriers, clinical realities further complicate treatment. Most current endovascular approaches, particularly flow diversion and stent-assisted coiling, require long-term dual antiplatelet therapy to prevent device-related thrombosis. While this treatment plan is necessary to reduce the risk of acute stent occlusion, it exposes patients to an ongoing danger of bleeding. Hemorrhagic complications have been reported in up to 8\% of patients \cite{ref40,ref41}, and the need for continuous blood clot inhibition complicates care when urgent surgery or traumatic injury occurs. These drawbacks make long-term management difficult and emphasize that, despite major advances, no existing therapy offers both durable occlusion and consistent safety across different aneurysm types.
\section*{Why This Problem is Critical}

The limitations of established therapies are especially concerning because of the profound consequences of rupture. Although SAH from ruptured aneurysms accounts for only about 10\% of all strokes, it remains one of the deadliest types. Mortality approaches 35\% within the first month, and nearly half of survivors live with long-term neurological disability \cite{ref1, ref3, ref42,ref43,ref44}. Despite advances in neurocritical care, rebleeding, vasospasm, and delayed ischemia continue to drive poor outcomes \cite{ref45}. Even survivors often experience cognitive and functional decline that limits independence and quality of life \cite{ref46}.

The challenge is amplified by the fact that not all aneurysms respond equally to current treatments. Wide-neck, fusiform, giant, and posterior circulation lesions consistently show lower rates of durable occlusion, even after technically successful interventions \cite{ref47}. These aneurysms remain at risk of recurrence and rupture, requiring patients to undergo repeated surveillance and retreatment. Aging populations further complicate the problem. Prevalence of unruptured aneurysms rises steadily with age, reaching nearly 6\% in individuals over 70. Older patients face higher rupture risks and tolerate invasive procedures, prolonged recovery, and dual antiplatelet therapy poorly \cite{ref48}.

Recognizing these gaps, recent years have seen the introduction of several device innovations. One major direction has been the development of surface-modified flow diverters, such as phosphorylcholine-coated devices like the Pipeline Shield (Medtronic) and hydrophilic-coated devices such as the p48MW HPC and p64MW HPC (Phenox). These coatings reduce platelet adhesion and appear to lower thromboembolic complication rates in registries \cite{ref34,ref49}. However, they cannot solve deployment challenges. Optical coherence tomography and Vaso-CT studies show that malapposition is common in tortuous arteries, with gaps of several millimeters between the device and the vessel wall. Such incomplete wall contact delays endothelial growth and permits persistent inflow, especially in fusiform and large aneurysms where thrombosis and occlusion are already slower \cite{ref31,ref32,ref50}. Thus, coatings improve hemocompatibility but do not address the geometric barriers that limit long-term success.

Another innovation has been the development of bioactive coils, which incorporate hydrogels, polymers, or drug-eluting coatings to encourage more complete aneurysm healing. The goal of these designs is to increase packing density inside the aneurysm sac and to stimulate a more stable fibrotic and endothelial response compared with bare platinum coils. For example, HydroCoils are coated with a hydrogel that swells once deployed, theoretically filling more of the aneurysm lumen and reducing empty space where recanalization can occur. The HELPS trial (HydroCoil Endovascular Aneurysm Occlusion and Packing Study) confirmed that these devices achieved slightly higher rates of stable occlusion compared with bare platinum coils \cite{ref53}. Despite these advances, limitations remain clear. Recurrence and retreatment are still frequent, especially in large and wide-necked aneurysms where the forces of blood flow and coil compaction are greatest \cite{ref54}. In such cases, even swelling polymers cannot provide the rigid, permanent barrier required to prevent coil migration or aneurysm reopening. Other bioactive designs, such as Matrix coils (coated with polyglycolic/polylactic acid polymers) and drug-eluting coils (designed to release antiproliferative or pro-healing factors), have shown some improvements in tissue response, but clinical trials and follow-up registries consistently report recanalization rates above 10–20\%. Meta-analyses of multiple coil platforms confirm that while bioactive coatings may delay the time to recurrence, they do not eliminate it, and many patients still require secondary treatments \cite{ref55}.

A third line of device innovation has focused on advancing current stents and developing bioresorbable scaffolds. Shape-memory stents, fabricated from thin-film nitinol (TFN), exploit the superelasticity of nitinol alloys to self-expand and adapt to tortuous vascular anatomy, thereby improving wall apposition compared with conventional braided metallic designs. Preclinical animal studies have reported favorable flexibility and partial endothelialization for these prototypes \cite{ref51,ref52}. In parallel, bioresorbable scaffolds such as the Esprit BTK (Abbott Vascular) provide temporary mechanical support and gradually resorb, aiming to overcome the long-term drawbacks of permanent metallic meshes. These platforms are designed to enhance hemocompatibility, conformability, and degradability with the ultimate goal of achieving safer and more durable aneurysm exclusion \cite{ref56,ref57,ref58}. Despite these advances, each approach faces significant limitations. Shape-memory nitinol stents remain at the preclinical stage, and although their superelasticity improves conformability, their mechanical strength is still lower than conventional metal stents, raising concerns about durability under pulsatile arterial flow. Deployment can also be unpredictable in tortuous or calcified vessels, as expansion depends on external triggers such as heat or light; variable local conditions may result in uneven expansion or incomplete wall contact.\cite{ref51,ref52}. Regarding bioresorbable scaffolds, they can deform, migrate, or even leak into the parent vessel during delivery, raising the risk of downstream embolization or unintended occlusion \cite{ref57}. Their mechanical strength also remains modest compared with metallic devices, casting doubt on their ability to withstand arterial pressures in wide-neck or high-flow aneurysms. Moreover, long-term outcomes for bioresorbable scaffolds remain uncertain, with limited evidence on degradation kinetics, inflammatory response, and reliable vessel wall integration \cite{ref55}. Most clinical reports are confined to small case series or short follow-up, leaving the durability of these technologies in complex aneurysm morphologies unresolved \cite{ref56,ref57,ref58}.

\begin{figure*}[ht]
\centering
\includegraphics[width=\textwidth]{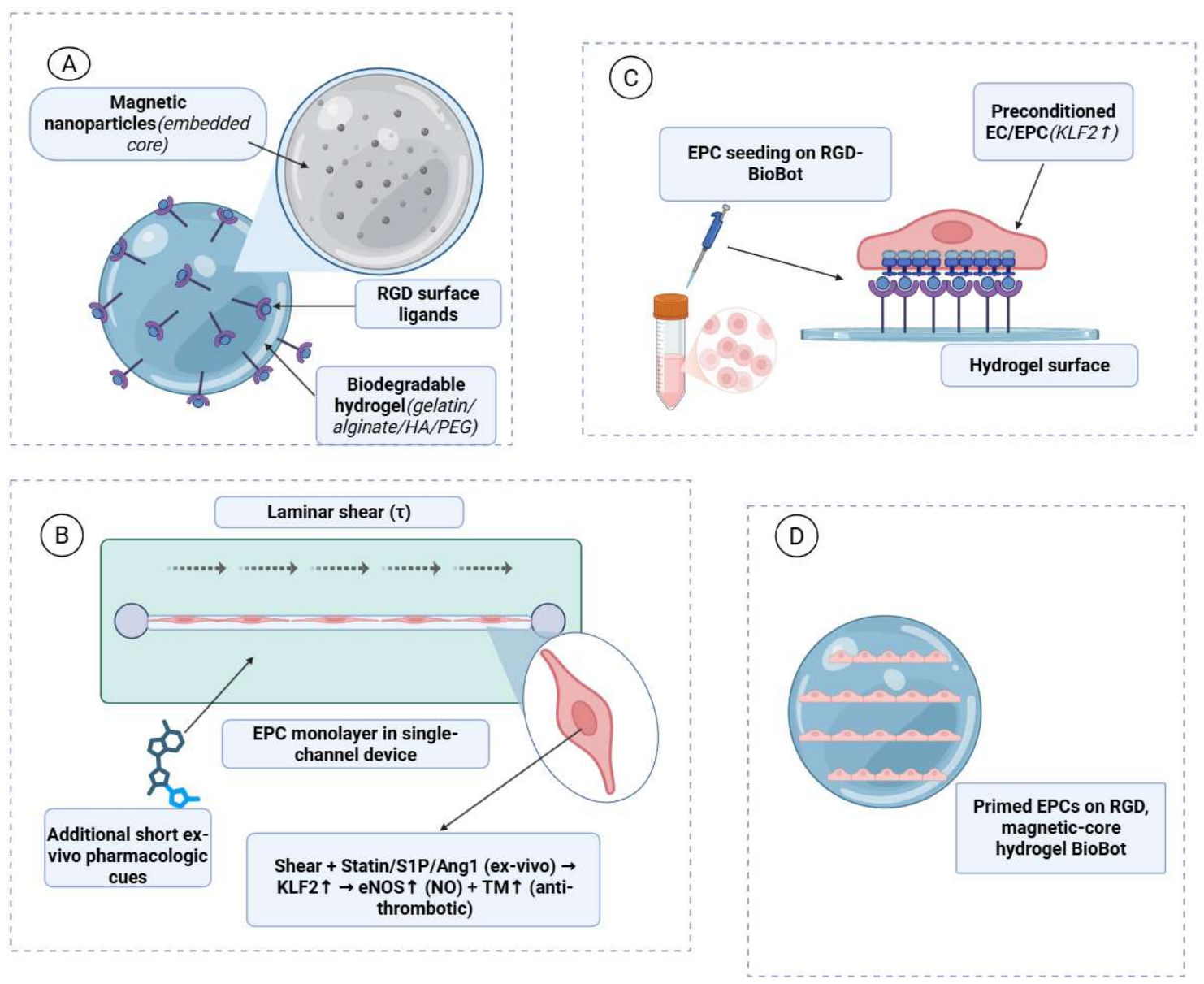}
\caption{ Design and preparation of BioBots. }
\label{fig:fullwidth}
\end{figure*} 
\section*{Introducing Magnetically Guided Endothelial BioBots}

Magnetically Guided Endothelial BioBots are envisioned as a next-generation therapy for cerebral aneurysms. Although coiling, clipping, and flow diversion are effective at modifying hemodynamics, they fail to restore the essential biological protection at the aneurysm neck, where the absence of an endothelial lining leaves the wall vulnerable. BioBots address this gap by combining magnetic microrobotics with endothelial biology.  Each BioBot is a biodegradable hydrogel, approximately 50–100 µm in diameter, embedded with magnetic nanoparticles and coated with primed endothelial cells (ECs) or endothelial progenitor cells (EPCs). This size range is small enough to pass through standard neurointerventional microcatheters yet large enough to support clusters of endothelial progenitor cells, balancing delivery feasibility with biological function.

The core architecture of BioBots is based on well-established cardiovascular biomaterials. Hydrogels such as gelatin, alginate, hyaluronic acid, and PEG derivatives are not chosen arbitrarily; they have been studied for decades in vascular repair because of their unique combination of safety, adaptability, and biological support \cite{ref60}. These polymers are biocompatible, meaning they do not trigger harmful immune or inflammatory responses when introduced into the body, and they can be engineered with precise control over stiffness, porosity, and degradation time. This flexibility is critical: a material that is too stiff would not integrate with fragile cerebral vessels, while one that degrades too quickly would not provide sufficient early support. To enable navigation, the hydrogel matrix is embedded with magnetic nanoparticles, which respond predictably to external electromagnetic fields. By combining the extracellular matrix–replicated properties of hydrogels with the magnetic sensitivity of magnetic cores, BioBots create a microenvironment that preserves endothelial viability under shear while also allowing precise positioning. This dual functionality ensures that cells adhere firmly, spread evenly, and function properly even in the dynamic setting of blood flow, while the carrier itself can be guided and compacted at the aneurysm neck.

To further improve adhesion, the carrier surfaces are functionalized with short peptide motifs such as RGD. These motifs serve as molecular anchors that bind to endothelial integrins, stabilizing attachment and promoting spreading \cite{ref61}. This interaction not only anchors the cells but also triggers intracellular signaling that supports survival, alignment, and barrier formation. As a result, the BioBots are not just inert scaffolds but actively engage the cells they carry. Recent advances in microfluidic fabrication enable these carriers to be made with high uniformity, each loaded with cells at the proper density and size \cite{ref62}. Crucially, their small diameter allows them to be injected through microcatheters used in routine neurointervention, bridging cutting-edge bioengineering with practical clinical delivery.

In addition, magnetic navigation provides precise control of BioBots inside the body. Embedded magnetic nanoparticles respond predictably to external electromagnetic fields, allowing steering without onboard motors or energy sources. This feature is particularly important in the brain, where tortuous vascular geometries make self-propelling micromachines impractical. Magnetic microrobotics has progressed significantly, and the principles of actuation, imaging, and control are now well defined \cite{ref63}. Proof-of-concept studies in animals have demonstrated that magnetic swarms can be concentrated at thrombus sites and used to mechanically enhance fibrinolysis under physiologic flow \cite{ref64}. Other platforms have successfully delivered therapeutic cargo to endovascular targets, showing that navigation and controlled deposition are feasible even under circulating conditions \cite{ref65}. Beyond laboratory testing, clinical-scale electromagnetic navigation arrays capable of generating programmable magnetic fields across human workspaces have been deployed and integrated into environments comparable to interventional radiology or neurosurgery suites \cite{ref66}.

For aneurysm therapy, BioBots could be released proximally through a microcatheter and then steered by external magnetic gradients directly to the aneurysm neck. Static or slowly varying fields would press the carriers laterally against the aneurysm margins, where they self-assemble into a conformal patch bridging the opening. This adaptability to irregular anatomy distinguishes them from rigid metallic stents, which often leave uncovered gaps in bifurcations, fusiform aneurysms, or highly curved arteries that sustain inflow and delay healing. BioBots, being soft and magnetically responsive, compact into virtually any geometry the aneurysm presents. Experimental data support this mechanism: magnetic swarms can be concentrated against vessel walls under flow \cite{ref67}, and magneto-responsive hydrogels can deform and align in response to applied fields \cite{ref68}. Once assembled, gentle oscillatory fields further reinforce the patch, while physiological shear stress orients endothelial cells along the vessel axis, ensuring both durable mechanical coverage and functional endothelial integration.

Equally important, the therapeutic effect of BioBots derives from ECs or EPCs that coat them.
\begin{figure*}[ht]
\centering
\includegraphics[width=\textwidth]{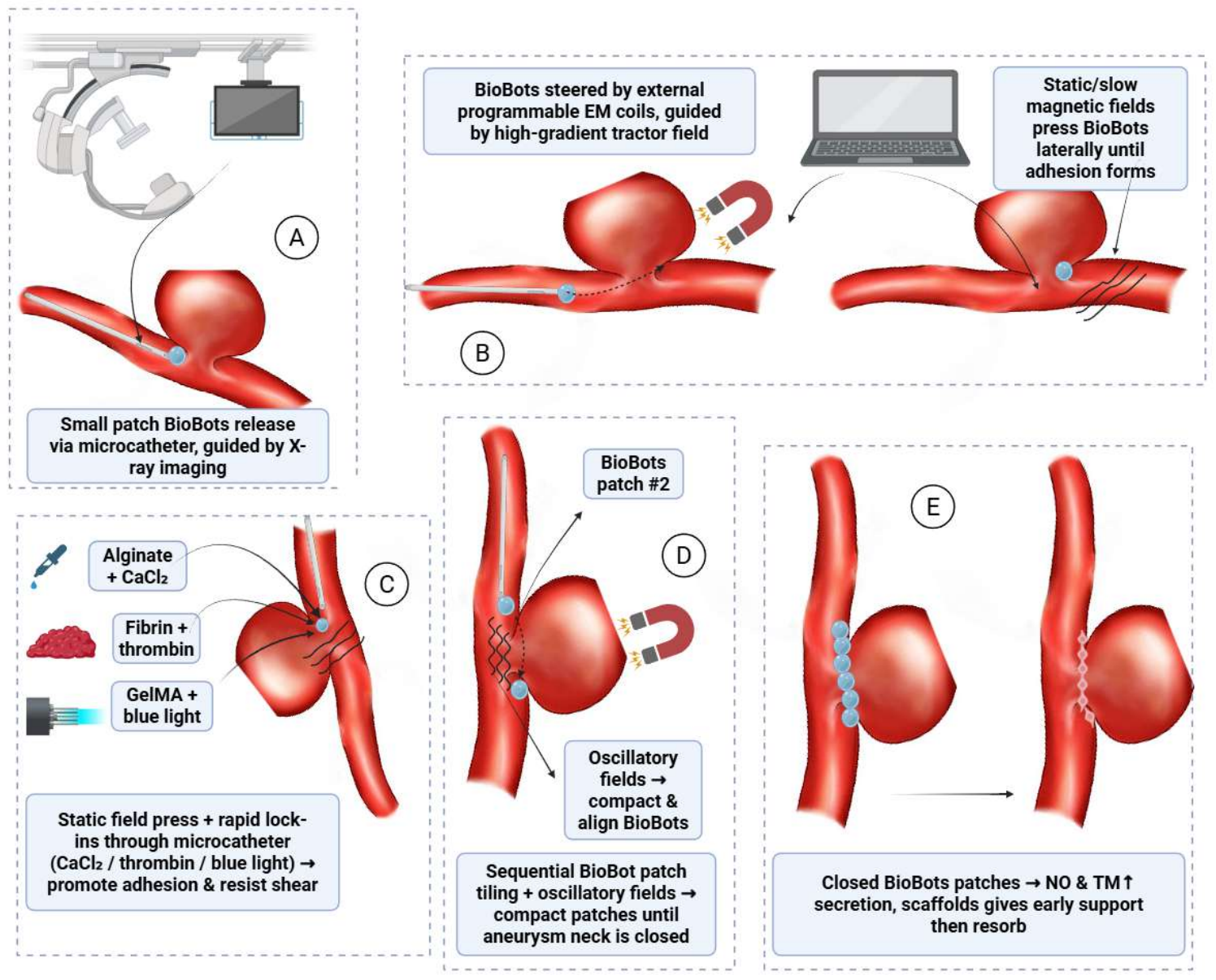} 
\caption{Therapeutic sequence of BioBots for aneurysm repair.}
\label{fig:fullwidth}
\end{figure*}
Before deployment, these cells are preconditioned under shear stress surrogates to activate the KLF2 transcriptional program, which upregulates protective molecules such as endothelial nitric oxide synthase (eNOS) and thrombomodulin \cite{ref69}. Nitric oxide generated by eNOS relaxes vessels, prevents vasospasm, and suppresses platelet aggregation, while thrombomodulin binds thrombin to limit clotting and inflammatory cascades \cite{ref70}. By delivering cells already in this vasoprotective state, BioBots provide an immediately functional endothelial lining at the aneurysm neck, bypassing the slow and variable process of natural endothelialization. Once placed in vivo, vascular shear reinforces this phenotype, maintaining cells in a stable, aligned, and quiescent state \cite{ref71}. Brief pharmacologic cues can further enhance this protective program: statins increase KLF2 and eNOS, sphingosine-1-phosphate (S1P) supports barrier integrity, and Ang1 or COMP-Ang1 stabilize junctions through Tie2 signaling, boosting thrombomodulin expression \cite{ref72,ref73,ref74}. Applied transiently during preconditioning and rinsed prior to seeding, these cues synergize with shear stress to lock cells into an antithrombotic, anti-inflammatory state, ensuring BioBots deliver a stably primed endothelium.

The choice of EPCs is further supported by real-world cardiovascular evidence. EPC-capture stents were developed to accelerate re-endothelialization of coronary implants. Clinical studies demonstrated that they did in fact lower thrombosis risk and speed endothelial coverage \cite{ref75}. Follow-up trials confirmed that EPCs reliably form stable linings on foreign surfaces, reducing adverse events such as restenosis \cite{ref76,ref77}. BioBots move beyond passive capture by delivering a concentrated, preconditioned EPC population directly to the site of vascular injury. This approach overcomes interpatient variability in circulating progenitor levels and provides a consistent basis for biological repair.

Additionally, to support these cells during the critical early phase after deployment, the hydrogel scaffold functions as transitional support. Immediately after deployment, the carriers bear hemodynamic stress and stabilize the patch against shear. During this early window, they protect the endothelial layer while cell–cell junctions mature. As the monolayer consolidates, the material degrades gradually, leaving only the autologous endothelium bridging the aneurysm neck. Hydrogels are particularly suited for this because they can be engineered for controlled resorption and have already been used to encapsulate endothelial cells or deliver extracellular vesicles in vascular contexts \cite{ref78}. When formulated as magnetically responsive composites, they can also temporarily stiffen under applied fields during assembly, then soften again once guidance is complete \cite{ref79}. This allows the therapeutic process to follow a predictable arc: delivery, navigation, assembly, stabilization, and scaffold resorption.

Therefore, the whole therapeutic sequence of BioBots can be described as a controlled timeline of delivery and biological integration. The process begins when a microcatheter is positioned just proximal to the aneurysm neck, through which BioBots are delivered in small patches. Each patch is guided by a programmable electromagnetic array; for example, clinical-scale systems such as the Stereotaxis Niobe or multi-coil electromagnetic arrays can generate a high-gradient “tractor” field that pulls the carriers toward the aneurysm defect, while a lower static field holds previously positioned patches securely against the vessel wall \cite{ref66}. Once localized, rapid anchoring mechanisms ensure resistance to washout under arterial shear. Immediate adhesion occurs through RGD–integrin interactions, mimicking natural extracellular matrix binding. Additional lock-in strategies may also be applied, including alginate shells that gel upon exposure to CaCl\textsubscript{2}, fibrin seals generated by microdoses of thrombin, or photo-crosslinkable GelMA hydrogels cured within minutes by catheter-mounted blue light \cite{ref80,ref81,ref82}. Magneto-responsive hydrogels further enhance stability, transiently stiffening under static magnetic fields to press BioBots firmly into the aneurysm margin until biological adhesion and crosslinking are complete. With successive releases, the patches tile into a conformal layer that adapts to irregular geometries such as bifurcation or fusiform aneurysms. From the earliest stage, the endothelial cells on the BioBots are already functional: preconditioned under laminar shear in vitro, they immediately secrete nitric oxide and thrombomodulin, creating a biologically active surface upon deployment. Over the following days, the cell layer matures into a continuous endothelial monolayer aligned with physiological shear. Tight and adherens junctions (claudins, occludin, and VE-cadherin) strengthen contacts between neighboring cells and improve barrier integrity. At the same time, the endothelium lays down its own extracellular matrix composed of laminin, collagen IV, nidogen, and fibronectin, forming a basement membrane beneath the cells. As this matrix accumulates it gradually takes over the mechanical and barrier roles of the scaffold. The hydrogel then resorbs on its programmed timeline, but coverage and function are preserved because the cells now rest on their own matrix. The end result is a stable, autologous endothelial lining that matches native vessel wall in structure and behavior 
\cite{ref60,ref78,ref83}.

\section*{Broader Real-World Evidence Guiding BioBot Design}

The plausibility of the BioBot concept is strongly reinforced by progress across multiple biomedical domains, where microrobotics and biohybrid systems have already demonstrated therapeutic success. These fields collectively show that the underlying principles of BioBots, including magnetic navigation, collective assembly under flow, biohybrid integration with living cells, and functional therapeutic activity, are achievable and clinically relevant.

In oncology, microrobots have been designed to transport chemotherapy into solid tumors. Traditional systemic chemotherapy often struggles with poor penetration into dense tumor tissue, leaving hypoxic cores untreated. Microrobotic platforms overcome this limitation by actively navigating through tissue microenvironments to deliver their cargo directly where it is needed. For example, engineered bacterial and algae-based microrobots have carried cytotoxic agents into tumors with high precision, achieving significantly deeper penetration and greater local retention compared with passive infusion \cite{ref84}. Building on this foundation, algae-based biohybrid microrobots have been used to deliver doxorubicin directly into metastatic lung tumors in vivo. These biohybrids not only reached the tumors but actively distributed drugs throughout the metastatic nodules, yielding stronger therapeutic effects than systemic delivery \cite{ref85}. These advances validate the principle that magnetically guided, biohybrid carriers can precisely navigate therapeutic cargo into anatomically challenging regions, a problem analogous to introducing endothelial cells across irregular aneurysm necks.

In vascular disease, the principle of collective magnetic navigation has already been tested in the context of clot dissolution. Swarming nanorobots, actuated by external magnetic fields, have been guided to thrombi inside vessels. Once localized, their coordinated mechanical motion disrupts fibrin networks and enhances fibrinolysis, significantly accelerating clot breakdown compared to drug alone \cite{ref64,ref65}. This provides direct proof that magnetic collectives can be concentrated in flowing blood, resist washout, and produce clinically relevant biological effects. These same features, including navigation, clustering under shear, and local therapeutic action, are central to BioBots’ proposed function at the aneurysm neck.

In infectious disease, microrobots have been deployed to penetrate bacterial biofilms, which represent one of the most formidable barriers to antibiotic therapy. Biofilms consist of dense extracellular polymeric matrices that protect bacteria from drug penetration. Magnetic microrobots have been shown to physically penetrate these barriers, disrupt the matrix, and release antimicrobials directly inside the biofilm \cite{ref86}. This demonstrates that microrobots can operate in dense, complex, and hostile microenvironments, maintaining functionality under mechanical and biochemical stress. The cerebral aneurysm environment, with its hemodynamic forces and irregular geometries, presents similar challenges that BioBots are designed to overcome.

In regenerative medicine, biohybrid microrobots that integrate living cells with responsive scaffolds have already been tested for tissue repair. For example, stem cell–laden magnetic carriers have been guided to damaged tissue sites, where they achieved higher engraftment efficiency and survival compared to unguided cell infusions \cite{ref87}. These findings confirm that combining cells with magnetically controllable scaffolds improves delivery, retention, and therapeutic effect, a principle that BioBots build upon by coupling preconditioned endothelial cells with biodegradable magnetic carriers for vascular healing.

\section*{How BioBots Could Treat Aneurysms Safely}

Magnetically Guided Endothelial BioBots are designed with safety integrated at multiple levels, beginning with the choice of scaffold materials. The carriers are constructed from biodegradable hydrogels and polymers such as gelatin, alginate, hyaluronic acid, and PEG derivatives, materials that have been used safely for decades in vascular and cardiac applications. These soft biomaterials degrade by hydrolysis or enzymatic action into nontoxic byproducts cleared naturally from the body. Unlike rigid metallic scaffolds, which can provoke chronic irritation or neointimal hyperplasia, the hydrogels conform to vascular anatomy without triggering foreign body responses. Their mechanical properties are adjustable to provide early protection against shear while avoiding excessive stress on fragile aneurysm walls, thereby lowering the likelihood of delayed inflammation or chronic remodeling.

Beyond the scaffold, safety is reinforced at the cellular level. The endothelial progenitor cells or endothelial cells coating the BioBots are not delivered in an unstable state but are preconditioned under laminar shear to activate KLF2. This preconditioning strongly upregulates eNOS and thrombomodulin, producing a phenotype that is anti-inflammatory, antithrombotic, and metabolically quiescent. By ensuring that the cells arrive already stabilized, BioBots minimize the risks of uncontrolled proliferation, thrombosis, or inflammatory activation. Importantly, once implanted, the natural shear environment of the cerebral arteries sustains this protective program, reducing the chance of phenotypic drift.

Safety also depends on where the carriers travel, and here magnetic navigation provides critical precision. One of the greatest concerns in microrobotics is off-target migration, but modern electromagnetic arrays can generate millimeter-scale field gradients that allow fine control of magnetic devices within tortuous vascular networks \cite{ref63,ref66}. In practice, BioBots would be introduced through a microcatheter placed just upstream of the aneurysm, minimizing dispersion, and then guided directly across the aneurysm neck one by one under fluoroscopic or MR imaging. Oscillatory or pulsed magnetic sequences can further compact the carriers into a dense endothelial patch, limiting the chance of stray particles moving downstream. Such precision has already been demonstrated in vivo for clot targeting and therapeutic delivery, providing realistic support for intracranial application \cite{ref64,ref65}.

These features distinguish BioBots from metallic implants. Flow diverters and stents rely on permanent meshes, which carry well-recognized complications such as malapposition in tortuous vessels, incomplete endothelialization, delayed thrombosis requiring long-term DAPT, and hemorrhagic events related to DAPT. Because BioBots leave no permanent scaffold, only a native endothelial lining remains once the hydrogel degrades. By avoiding metallic struts, this approach could substantially reduce or even eliminate the need for prolonged DAPT, offering a major advantage for patients at elevated risk of hemorrhage.
Finally, degradability adds additional safety. If a subset of carriers fails to assemble, migrates off target, or lodges in an unintended location, they degrade naturally over days to weeks into inert byproducts. This built-in temporality ensures that procedural imperfections do not translate into permanent harm, unlike misplaced metallic implants that persist indefinitely. In this way, BioBots provide therapeutic benefit while minimizing long-term risk, uniting multiple layers of safety in a single platform.

\section*{Research Roadmap}
The successful translation of Magnetically Guided Endothelial BioBots from concept to clinical therapy requires a staged research program that advances step by step from controlled laboratory systems to animal validation and ultimately to translational development. This pathway integrates both biological and engineering perspectives, ensuring that each stage builds directly on the evidence of the one before it.

The process begins with in vitro validation, where BioBots are tested in controlled aneurysm-on-chip systems. Microfluidic or PDMS-based aneurysm phantoms can be fabricated with physiologically relevant geometries and perfused with blood-mimicking fluids to reproduce hemodynamic shear patterns. Within these models, BioBot navigation can be evaluated under flow, assessing how external magnetic fields guide and concentrate them across aneurysm necks. High-resolution particle image velocimetry (PIV) and computational fluid dynamics (CFD) analysis can then quantify changes in inflow, wall shear stress, and vorticity once BioBots assemble. At the same time, endothelial viability, orientation, and phenotype stability can be monitored over days to confirm that preconditioned EPCs maintain KLF2 and eNOS expression under dynamic shear. These early studies provide a safe, highly controllable environment to establish proof of principle.

Building on in vitro findings, the next stage moves into in vivo studies using established animal aneurysm models such as elastase-induced rabbit carotid aneurysms, canine bifurcation aneurysms, or swine sidewall aneurysms. These models make it possible to test BioBots under the full complexity of living vascular systems that mimic wide-neck and fusiform lesions. Key endpoints include successful navigation and assembly under physiologic blood flow, durability of endothelial patch retention, maturation of the endothelial lining, kinetics of scaffold degradation, and safety outcomes such as absence of distal emboli, inflammatory responses, or systemic toxicity. Serial imaging (angiography, OCT, MRI) combined with histology at defined intervals provides both structural and biological markers, delivering the first in vivo evidence of performance and safety.

Once safety and efficacy are demonstrated in animals, the next stage focuses on translational development, adapting BioBots for clinical workflows and meeting regulatory safety standards. Catheter-based delivery must be integrated with clinically scaled electromagnetic navigation arrays capable of generating field strengths and gradients across the human cranial workspace. This phase proceeds stepwise: first, bench-top integration of catheters and magnets in anatomically realistic phantoms; next, pilot navigation studies in large animals under interventional imaging; and finally, comprehensive safety validation under good laboratory practice (GLP). Endpoints include thrombogenicity, immunogenicity, biodistribution of degradation products, and long-term vessel patency. Together, these translational studies form the bridge between preclinical success and a first-in-human feasibility trial.
\section*{Acknowledging Limitations}
Although the BioBot concept builds on advances in biomaterials, microrobotics, and vascular biology, it remains theoretical and faces significant challenges. A major obstacle is navigation and deployment within the cerebral vasculature. Intracranial arteries are among the most demanding sites in the body, with small calibers, tortuous anatomy, and constant exposure to pulsatile high-shear flow. Current electromagnetic arrays have demonstrated millimeter-scale precision, but finer control is required to achieve the submillimeter targeting necessary for safe use in these vessels. Importantly, while regenerative microrobot studies have already shown success in peripheral tissues such as skeletal muscle, ischemic myocardium, and dermal wounds, these environments are far more accessible and geometrically less complex than the cerebral circulation. No biohybrid microrobot system has yet been tested in the brain, underscoring both the novelty of the BioBot concept and the magnitude of the translational challenge.

On the biological side, while shear preconditioning activates protective KLF2 and eNOS pathways, the long-term survival and integration of endothelial cells in the intracranial environment, with its high shear, inflammatory mediators, and patient-to-patient variability, are not assured. Another uncertainty lies in scaffold resorption. If degradation occurs too quickly, BioBots may lose structural support before endothelial junctions and extracellular matrix deposition stabilize the patch. Conversely, overly slow degradation risks persistent material, inflammatory by-products, and interference with vessel remodeling. Careful tuning of hydrogel chemistry and crosslinking will therefore be critical to align degradation kinetics with the timeframe of endothelial maturation and basement membrane formation.

Safety questions also remain. Misdirected carriers could embolize distal branches, and although hydrogels are generally biocompatible, their degradation products and embedded nanoparticles require rigorous evaluation for immune, thrombotic, and inflammatory effects. Imaging and intraoperative monitoring pose further difficulties: fluoroscopy can visualize metallic particles, but noninvasive tracking of soft hydrogel carriers is limited, and MRI compatibility with magnetic nanoparticles must be validated.

Beyond technical and biological concerns, translational barriers remain substantial. Endothelial progenitor cells vary widely between donors, complicating the development of reliable, large-scale sources for therapy. Manufacturing adds another hurdle, since producing magnetically responsive, cell-laden hydrogels at clinical scale will demand rigorous quality control and compliance with Good Manufacturing Practices. Regulatory classification is also complex, as BioBots span the boundaries of device and biologic regulation, requiring stepwise validation and coordinated oversight. Finally, practical and economic considerations cannot be overlooked: compared with existing coils and flow diverters, BioBot therapies may initially be more complex and costly, slowing adoption unless they demonstrate clear advantages in safety and long-term outcomes.

\section*{Future Outlook \& Conclusion}

Magnetically Guided Endothelial BioBots suggest a shift in aneurysm treatment from mechanical occlusion toward living vascular reconstruction. By combining microrobotic navigation with endothelial biology, this approach directly addresses many of the current limitations of clipping, coiling, and flow diversion, including poor conformability, delayed healing, and reliance on rigid metallic implants. These benefits could be especially impactful in complex aneurysms such as wide-neck, giant, fusiform, or bifurcated lesions, where existing devices frequently fail to provide safe and durable repair. Moving forward, collaboration between engineers, biologists, and clinicians will be essential to refine navigation systems, optimize cell–scaffold interactions, and resolve regulatory hurdles. If these challenges can be overcome, BioBots may provide a transformative treatment that not only overcomes the shortcomings of current methods but also delivers particular advantages for the most difficult aneurysms to treat.

\section*{Declarations}

\subsection*{Conflict of Interest}
The author declares no conflicts of interest.

\subsection*{Ethical Approval}
Not applicable. This review did not involve studies with human participants or animals.

\subsection*{Funding}
This work received no external funding.

\subsection*{Author Contributions}
Duong Le was responsible for conceptualization, original draft preparation, and subsequent review and editing of the manuscript.

\subsection*{Data Availability}
Not applicable. No new data were created or analyzed in this review.

\clearpage            


\begin{thebibliography}{87}

\bibitem{ref1}
Deshmukh AS, Priola SM, Katsanos AH, Scalia G, Costa Alves A, Srivastava A, et al.
The management of intracranial aneurysms: current trends and future directions.
\textit{Neurol Int}. 2024;16(1):74--94.
\href{https://doi.org/10.3390/neurolint16010005}{doi:10.3390/neurolint16010005}.

\bibitem{ref2}
Ortiz AFH, Suriano ES, Eltawil Y, Sekhon M, Gebran A, Garland M, et al.
Prevalence and risk factors of unruptured intracranial aneurysms in ischemic stroke patients - a global meta-analysis.
\textit{Surg Neurol Int}. 2023;14:222.
\href{https://doi.org/10.25259/SNI_287_2023}{doi:10.25259/SNI\_287\_2023}


\bibitem{ref3}
Medscape. Cerebral aneurysm: background, epidemiology, etiology [Internet].
New York (NY): WebMD LLC; [cited 2025 August 21].
Available from: \href{https://emedicine.medscape.com/article/1161518-overview}{https://emedicine.medscape.com/article/1161518-overview}.

\bibitem{ref4}
Kawashima M, Rhoton AL Jr, Tanriover N, Ulm AJ, Yasuda A, Fujii K.
Fusiform aneurysms: a review from its pathogenesis to treatment options.
\textit{Surg Neurol Int}, 2021;12:471.
\href{https://doi.org/10.25259/SNI_658_2021}{\texttt{doi:10.25259/SNI\_658\_2021}}


\bibitem{ref5}
Hendricks BK, Yoon JS, Yaeger K, et al.
Wide-neck aneurysms: systematic review of the neurosurgical literature with a focus on definition and clinical implications.
\textit{J Neurosurg}, 2019;133(1):159--165.
\href{https://doi.org/10.3171/2019.3.JNS183160}{doi:10.3171/2019.3.JNS183160}.

\bibitem{ref6}
Belavadi R, Gudigopuram SVR, Raguthu CC, Gajjela H, Kela I, Kakarala CL, et al.
Surgical clipping versus endovascular coiling in the management of intracranial aneurysms.
\textit{Cureus}. 2021;13(12):e20478.
\href{https://doi.org/10.7759/cureus.20478}{doi:10.7759/cureus.20478}.

\bibitem{ref7}
Andereggen L, Bosshart SL, Marbacher S, Grüter BE, Berberat J, Schubert GA, et al.
Long-term hemorrhage and reperfusion rates of coiled aneurysms: a single-center experience.
\textit{J Clin Med}. 2024;13(17):5223.
\href{https://doi.org/10.3390/jcm13175223}{doi:10.3390/jcm13175223}.

\bibitem{ref8}
Johnston SC, Dowd CF, Higashida RT, Lawton MT, Duckwiler GR, Gress DR, et al.
Predictors of rehemorrhage after treatment of ruptured intracranial aneurysms: the Cerebral Aneurysm Rerupture After Treatment (CARAT) study.
\textit{Stroke}. 2008;39(1):120--5.
\href{https://doi.org/10.1161/STROKEAHA.107.495747}{doi:10.1161/STROKEAHA.107.495747}.

\bibitem{ref9}
Kannath SK, Mohimen A, Raman KT, Abraham M, Nair S, Rajan JE.
Single centre experience of flow diverter treatment of complex intracranial aneurysms from South India: intermediate and long-term outcomes.
\textit{Neurol India}. 2019;67(3):797--802.
\href{https://doi.org/10.4103/0028-3886.263259}{doi:10.4103/0028-3886.263259}.

\bibitem{ref10}
Hohenstatt S, Arrichiello A, Conte G, Craparo G, Caranci F, Angileri A, et al.
Branch vessel occlusion in aneurysm treatment with flow diverter stent.
\textit{Acta Biomed}. 2020;91(10-S):e2020003.
\href{https://doi.org/10.23750/abm.v91i10-S.10317}{doi:10.23750/abm.v91i10-S.10317}.

\bibitem{ref11}
Liu J, Jing L, Wang C, Zhang Y, Yang X.
Recanalization, regrowth, and delayed rupture of a previously coiled unruptured anterior communicating artery aneurysm: a longitudinal hemodynamic analysis.
\textit{World Neurosurg}. 2016;89:726.e5--726.e10.
\href{https://doi.org/10.1016/j.wneu.2015.12.057}{doi:10.1016/j.wneu.2015.12.057}.

\bibitem{ref12}
Enriquez-Marulanda A, Young MM, Taussky P.
Flow diversion: a disruptive technology coming of age. Lessons learned and challenges for the future.
\textit{J Neurosurg}. 2023;139(5):1317--27.
\href{https://doi.org/10.3171/2023.3.JNS23167}{doi:10.3171/2023.3.JNS23167}.

\bibitem{ref13}
Oliver AA, Ognard J, Cortese J, Bayraktar EA, Zielonka AM, Kallmes DF, et al.
Magnetizable flow diverters can magnetically capture and retain endothelial cells to promote healing in rabbit arteries.
\textit{J Neuroradiol}. 2025;52(4):101353.
\href{https://doi.org/10.1016/j.neurad.2025.101353}{doi:10.1016/j.neurad.2025.101353}.

\bibitem{ref14}
Seibert B, Tummala RP, Chow R, Faridar A, Mousavi SA, Divani AA.
Intracranial aneurysms: review of current treatment options and outcomes.
\textit{Front Neurol}. 2011;2:45.
\href{https://doi.org/10.3389/fneur.2011.00045}{doi:10.3389/fneur.2011.00045}.

\bibitem{ref15}
Fotakopoulos G, Lempesis IG, Georgakopoulou VE, Trakas N, Sklapani P, Faropoulos K, et al.
Surgical outcomes of patients with unruptured anterior vs. inferior circulation aneurysms: a meta-analysis.
\textit{Med Int (Lond)}. 2023;4(1):5.
\href{https://doi.org/10.3892/mi.2023.129}{doi:10.3892/mi.2023.129}.

\bibitem{ref16}
Spetzler RF, McDougall CG, Zabramski JM, Albuquerque FC, Hills NK, Russin JJ, et al.
The Barrow Ruptured Aneurysm Trial: 6-year results.
\textit{J Neurosurg}. 2015;123(3):609--17.
\href{https://doi.org/10.3171/2014.9.JNS141749}{doi:10.3171/2014.9.JNS141749}.

\bibitem{ref17}
Wiebers DO, Whisnant JP, Huston J 3rd, Meissner I, Brown RD Jr, Piepgras DG, et al.
Unruptured intracranial aneurysms: natural history, clinical outcome, and risks of surgical and endovascular treatment.
\textit{Lancet}. 2003;362(9378):103--10.
\href{https://doi.org/10.1016/S0140-6736(03)13860-3}{doi:10.1016/S0140-6736(03)13860-3}.

\bibitem{ref18}
Lawton MT, Vates GE.
Subarachnoid hemorrhage.
\textit{N Engl J Med}. 2017;377(3):257--66.
\href{https://doi.org/10.1056/NEJMcp1605827}{doi:10.1056/NEJMcp1605827}.

\bibitem{ref19}
Barker FG 2nd, Amin-Hanjani S, Butler WE, Hoh BL, Rabinov JD, Pryor JC, et al.
Age-dependent differences in short-term outcome after surgical or endovascular treatment of unruptured intracranial aneurysms in the United States, 1996--2000.
\textit{Neurosurgery}. 2004;54(1):18--28; discussion 28--30.
\href{https://doi.org/10.1227/01.NEU.0000097195.83626.C5}{doi:10.1227/01.NEU.0000097195.83626.C5}.

\bibitem{ref20}
Raymond J, Guilbert F, Weill A, Georganos SA, Juravsky L, Lambert A, et al.
Long-term angiographic recurrences after selective endovascular treatment of aneurysms with detachable coils.
\textit{Stroke}. 2003;34(6):1398--403.
\href{https://doi.org/10.1161/01.STR.0000073841.88563.E9}{doi:10.1161/01.STR.0000073841.88563.E9}.

\bibitem{ref21}
Grunwald IQ, Papanagiotou P, Struffert T, Politi M, Krick C, Gül G, et al.
Recanalization after endovascular treatment of intracerebral aneurysms.
\textit{Neuroradiology}. 2007;49(1):41--7.
\href{https://doi.org/10.1007/s00234-006-0153-4}{doi:10.1007/s00234-006-0153-4}.

\bibitem{ref22}
Piotin M, Spelle L, Mounayer C, Salles-Rezende MT, Giansante-Abud D, Vanzin-Santos R, et al.
Intracranial aneurysms: treatment with bare platinum coils—aneurysm packing, complex coils, and angiographic recurrence.
\textit{Radiology}. 2007;243(2):500--8.
\href{https://doi.org/10.1148/radiol.2431060006}{doi:10.1148/radiol.2431060006}.

\bibitem{ref23}
Murayama Y, Nien YL, Duckwiler G, Gobin YP, Jahan R, Frazee J, et al.
Guglielmi detachable coil embolization of cerebral aneurysms: 11 years' experience.
\textit{J Neurosurg}. 2003;98(5):959--66.
\href{https://doi.org/10.3171/jns.2003.98.5.0959}{doi:10.3171/jns.2003.98.5.0959}.

\bibitem{ref24}
Zhang X, Zuo Q, Tang H, Xue G, Yang P, Zhao R, et al.
Stent assisted coiling versus non-stent assisted coiling for the management of ruptured intracranial aneurysms: a meta-analysis and systematic review.
\textit{J Neurointerv Surg}. 2019;11(5):489--96.
\href{https://doi.org/10.1136/neurintsurg-2018-014346}{doi:10.1136/neurintsurg-2018-014346}.

\bibitem{ref25}
Molyneux AJ, Kerr RS, Yu LM, Clarke M, Sneade M, Yarnold JA, et al.
International subarachnoid aneurysm trial (ISAT) of neurosurgical clipping versus endovascular coiling in 2143 patients with ruptured intracranial aneurysms: a randomised comparison of effects on survival, dependency, seizures, rebleeding, subgroups, and aneurysm occlusion.
\textit{Lancet}. 2005;366(9488):809--17.
\href{https://doi.org/10.1016/S0140-6736(05)67214-5}{doi:10.1016/S0140-6736(05)67214-5}.

\bibitem{ref26}
Molyneux AJ, Kerr RS, Birks J, Ramzi N, Yarnold J, Sneade M, et al.
Risk of recurrent subarachnoid haemorrhage, death, or dependence and standardised mortality ratios after clipping or coiling of an intracranial aneurysm in the International Subarachnoid Aneurysm Trial (ISAT): long-term follow-up.
\textit{Lancet Neurol}. 2009;8(5):427--33.
\href{https://doi.org/10.1016/S1474-4422(09)70080-8}{doi:10.1016/S1474-4422(09)70080-8}.

\bibitem{ref27}
Kallmes DF, Brinjikji W, Cekirge S, Fiorella D, Hanel RA, Jabbour P, et al.
Safety and efficacy of the Pipeline embolization device for treatment of intracranial aneurysms: a pooled analysis of 3 large studies.
\textit{J Neurosurg}. 2017;127(4):775--80.
\href{https://doi.org/10.3171/2016.8.JNS16467}{doi:10.3171/2016.8.JNS16467}.

\bibitem{ref28}
Shehata MA, Ibrahim MK, Ghozy S, Bilgin C, Jabal MS, Kadirvel R, et al.
Long-term outcomes of flow diversion for unruptured intracranial aneurysms: a systematic review and meta-analysis.
\textit{J Neurointerv Surg}. 2023;15(9):898--902.
\href{https://doi.org/10.1136/jnis-2022-019240}{doi:10.1136/jnis-2022-019240}.

\bibitem{ref29}
Rouchaud A, Brinjikji W, Lanzino G, Cloft HJ, Kadirvel R, Kallmes DF.
Delayed hemorrhagic complications after flow diversion for intracranial aneurysms: a literature overview.
\textit{Neuroradiology}. 2016;58(2):171--7.
\href{https://doi.org/10.1007/s00234-015-1615-4}{doi:10.1007/s00234-015-1615-4}.

\bibitem{ref30}
Kühn AL, Rodrigues KM, Wakhloo AK, Puri AS.
Endovascular techniques for achievement of better flow diverter wall apposition.
\textit{Interv Neuroradiol}. 2019;25(3):344--7.
\href{https://doi.org/10.1177/1591019918815224}{doi:10.1177/1591019918815224}.

\bibitem{ref31}
Wan Z, Liu T, Xu N, Zhu W, Qi Y, Ma C, et al.
Flow diverter tail malapposition after implantation in the internal carotid artery for aneurysm treatment: a preliminary study.
\textit{Front Neurol}. 2023;14:1301046.
\href{https://doi.org/10.3389/fneur.2023.1301046}{doi:10.3389/fneur.2023.1301046}.

\bibitem{ref32}
Becske T, Brinjikji W, Potts MB, Kallmes DF, Shapiro M, Moran CJ, et al.
Long-term clinical and angiographic outcomes following Pipeline Embolization Device treatment of complex internal carotid artery aneurysms: five-year results of the Pipeline for Uncoilable or Failed Aneurysms trial.
\textit{Neurosurgery}. 2017;80(1):40--8.
\href{https://doi.org/10.1093/neuros/nyw014}{doi:10.1093/neuros/nyw014}.

\bibitem{ref33}
Narayan D, Venkatraman SS.
Effect of pore size and interpore distance on endothelial cell growth on polymers.
\textit{J Biomed Mater Res A}. 2008;87(3):710--8.
\href{https://doi.org/10.1002/jbm.a.31824}{doi:10.1002/jbm.a.31824}.

\bibitem{ref34}
Salem MM, Sweid A, Kuhn AL, et al.
Repeat flow diversion for cerebral aneurysms failing prior flow diversion: safety and feasibility from multicenter experience.
\textit{Stroke}. 2022;53(4):1178--89.
\href{https://doi.org/10.1161/STROKEAHA.120.033555}{doi:10.1161/STROKEAHA.120.033555}.

\bibitem{ref35}
Sharashidze V, Raz E, Nossek E, Kvint S, Riina H, Rutledge C, et al.
Comprehensive analysis of post-Pipeline endothelialization and remodeling.
\textit{AJNR Am J Neuroradiol}. 2024;45(7):893--8.
\href{https://doi.org/10.3174/ajnr.A8246}{doi:10.3174/ajnr.A8246}.

\bibitem{ref36}
Frösen J, Tulamo R, Paetau A, Laaksamo E, Korja M, Laakso A, et al.
Saccular intracranial aneurysm: pathology and mechanisms.
\textit{Acta Neuropathol}. 2012;123(6):773--86.
\href{https://doi.org/10.1007/s00401-011-0939-3}{doi:10.1007/s00401-011-0939-3}.

\bibitem{ref37}
Chalouhi N, Ali MS, Jabbour PM, Tjoumakaris SI, Gonzalez LF, Rosenwasser RH, et al.
Biology of intracranial aneurysms: role of inflammation.
\textit{J Cereb Blood Flow Metab}. 2012;32(9):1659--76.
\href{https://doi.org/10.1038/jcbfm.2012.84}{doi:10.1038/jcbfm.2012.84}.

\bibitem{ref38}
Buyukkaya R, Kocaeli H, Yildirim N, Cebeci H, Erdogan C, Hakyemez B.
Treatment of complex intracranial aneurysms using flow-diverting silk® stents. An analysis of 32 consecutive patients.
\textit{Interv Neuroradiol}. 2014;20(6):729--35.
\href{https://doi.org/10.15274/INR-2014-10070}{doi:10.15274/INR-2014-10070}.

\bibitem{ref39}
Bahar A, Kohar RC, Gunawan A, Nurwati D, Ramadhany S, Widiastomo T, et al.
Single versus multiple coverage of Pipeline embolization device for treatment of intracranial aneurysms: a systematic review.
\textit{Egypt J Neurol Psychiatr Neurosurg}. 2023;59:130.
\href{https://doi.org/10.1186/s41983-023-00731-7}{doi:10.1186/s41983-023-00731-7}.

\bibitem{ref40}
Toma A, Essibayi MA, Osama M, Karandish A, Dmytriw AA, Altschul D.
Managing thrombosis risk in flow diversion: a review of antiplatelet approaches.
\textit{Neuroradiol J}. 2025:19714009251313515.
\href{https://doi.org/10.1177/19714009251313515}{doi:10.1177/19714009251313515}.

\bibitem{ref41}
Abo Kasem R, Hubbard Z, Cunningham C, Almorawed H, Isidor J, Samman Tahhan I, et al.
Comparison of flow diverter alone versus flow diverter with coiling for large and giant intracranial aneurysms: systematic review and meta-analysis of observational studies.
\textit{J Neurointerv Surg}. 2025.
\href{https://doi.org/10.1136/jnis-2024-022845}{doi:10.1136/jnis-2024-022845}.

\bibitem{ref42}
Kurisu K, Sakata H, Matsumoto Y, Kanoke A, Omodaka S, Fujimura M, et al.
Repeated recurrence after endovascular treatment for cerebral aneurysms: predictive clinical factors and optimal therapeutic management.
\textit{Neurosurg Rev}. 2025;48(1):603.
\href{https://doi.org/10.1007/s10143-024-02585-9}{doi:10.1007/s10143-024-02585-9}.

\bibitem{ref43}
de Rooij NK, Linn FH, van der Plas JA, Algra A, Rinkel GJ.
Incidence of subarachnoid haemorrhage: a systematic review with emphasis on region, age, gender and time trends.
\textit{J Neurol Neurosurg Psychiatry}. 2007;78(12):1365--72.
\href{https://doi.org/10.1136/jnnp.2007.117655}{doi:10.1136/jnnp.2007.117655}.

\bibitem{ref44}
Nieuwkamp DJ, Setz LE, Algra A, Linn FH, de Rooij NK, Rinkel GJ.
Changes in case fatality of aneurysmal subarachnoid haemorrhage over time, according to age, sex, and region: a meta-analysis.
\textit{Lancet Neurol}. 2009;8(7):635--42.
\href{https://doi.org/10.1016/S1474-4422(09)70126-7}{doi:10.1016/S1474-4422(09)70126-7}.

\bibitem{ref45}
Claassen J, Park S.
Spontaneous subarachnoid haemorrhage.
\textit{Lancet}. 2022;400(10355):846--62.
\href{https://doi.org/10.1016/S0140-6736(22)00938-2}{doi:10.1016/S0140-6736(22)00938-2}.

\bibitem{ref46}
Greebe P, Rinkel GJ, Hop JW, Visser-Meily JM, Algra A.
Functional outcome and quality of life 5 and 12.5 years after aneurysmal subarachnoid haemorrhage.
\textit{J Neurol}. 2010;257(12):2059--64.
\href{https://doi.org/10.1007/s00415-010-5660-2}{doi:10.1007/s00415-010-5660-2}.

\bibitem{ref47}
Brinjikji W, Murad MH, Lanzino G, Cloft HJ, Kallmes DF.
Endovascular treatment of intracranial aneurysms with flow diverters: a meta-analysis.
\textit{Stroke}. 2013;44(2):442--7.
\href{https://doi.org/10.1161/STROKEAHA.112.678151}{doi:10.1161/STROKEAHA.112.678151}.

\bibitem{ref48}
Vlak MH, Algra A, Brandenburg R, Rinkel GJ.
Prevalence of unruptured intracranial aneurysms, with emphasis on sex, age, comorbidity, country, and time period: a systematic review and meta-analysis.
\textit{Lancet Neurol}. 2011;10(7):626--36.
\href{https://doi.org/10.1016/S1474-4422(11)70109-0}{doi:10.1016/S1474-4422(11)70109-0}.

\bibitem{ref49}
Goertz L, Weyland CS, Nikoubashman O, Bürkle F, de Beukelaer F, Siebert E, et al.
Multicenter study of HPC coated p48 and p64 flow diverters for treatment of intracranial aneurysms under dual antiplatelet therapy.
\textit{Interv Neuroradiol}. 2025:15910199251318066.
\href{https://doi.org/10.1177/15910199251318066}{doi:10.1177/15910199251318066}.

\bibitem{ref50}
van der Marel K, Gounis MJ, Weaver JP, de Korte AM, King RM, Arends JM, et al.
Grading of regional apposition after flow-diverter treatment (GRAFT): a comparative evaluation of VasoCT and intravascular OCT.
\textit{J Neurointerv Surg}. 2016;8(8):847--52.
\href{https://doi.org/10.1136/neurintsurg-2015-011843}{doi:10.1136/neurintsurg-2015-011843}.

\bibitem{ref51}
Pineda-Castillo SA, Stiles AM, Bohnstedt BN, Lee H, Liu Y, Lee CH.
Shape memory polymer-based endovascular devices: design criteria and future perspective.
\textit{Polymers (Basel)}. 2022;14(13):2526.
\href{https://doi.org/10.3390/polym14132526}{doi:10.3390/polym14132526}.

\bibitem{ref52}
Zhao S, Gu L, Froemming SR.
Performance of self-expanding nitinol stent in a curved artery: impact of stent length and deployment orientation.
\textit{J Biomech Eng}. 2012;134(7):071007.
\href{https://doi.org/10.1115/1.4007095}{doi:10.1115/1.4007095}.

\bibitem{ref53}
White PM, Lewis SC, Gholkar A, Sellar RJ, Nahser H, Cognard C, et al.
Hydrogel-coated coils versus bare platinum coils for the endovascular treatment of intracranial aneurysms (HELPS): a randomised controlled trial.
\textit{Lancet}. 2011;377(9778):1655--62.
\href{https://doi.org/10.1016/S0140-6736(11)60408-X}{doi:10.1016/S0140-6736(11)60408-X}.

\bibitem{ref54}
McDougall CG, Johnston SC, Hetts SW, Gholkar A, Barnwell SL, Vazquez Suarez JC, et al.
Five-year results of randomized bioactive versus bare metal coils in the treatment of intracranial aneurysms: the Matrix and Platinum Science (MAPS) Trial.
\textit{J Neurointerv Surg}. 2021;13(10):930--4.
\href{https://doi.org/10.1136/neurintsurg-2020-016906}{doi:10.1136/neurintsurg-2020-016906}.

\bibitem{ref55}
Cortese J, Ghozy S, Zarrintan A, et al.
Hydrogel coils versus bare platinum coils for the treatment of ruptured and unruptured aneurysms: an updated systematic review and meta-analysis of randomized controlled trials.
\textit{AJNR Am J Neuroradiol}. 2025;46(7):1379--86.
\href{https://doi.org/10.3174/ajnr.A8697}{doi:10.3174/ajnr.A8697}.

\bibitem{ref56}
Poupart O, Schmocker A, Conti R, Meuli R, Poleni PE, Machi P, et al.
In vitro implementation of photopolymerizable hydrogels as a potential treatment of intracranial aneurysms.
\textit{Front Bioeng Biotechnol}. 2020;8:261.
\href{https://doi.org/10.3389/fbioe.2020.00261}{doi:10.3390/fbioe.2020.00261}.

\bibitem{ref57}
Kim S, Nowicki KW, Kohyama K, et al.
Development of an injectable, ECM-derivative embolic for the treatment of cerebral saccular aneurysms.
\textit{Bomacromolecules}. 2024;25(8):4879--90.
\href{https://doi.org/10.1021/acs.biomac.4c00321}{doi:10.1021/acs.biomac.4c00321}.

\bibitem{ref58}
Rotaru-Zăvăleanu AD, Dinescu VC, Aldea M, Gresita A.
Hydrogel-based therapies for ischemic and hemorrhagic stroke: a comprehensive review.
\textit{Gels}. 2024;10(7):476.
\href{https://doi.org/10.3390/gels10070476}{doi:10.3390/gels10070476}.

\bibitem{ref59}
Zhou H, Mayorga-Martinez CC, Pané S, Zhang L, Pumera M.
Magnetically driven micro and nanorobots.
\textit{Chem Rev}. 2021;121(8):4999--5041.
\href{https://doi.org/10.1021/acs.chemrev.0c01234}{doi:10.1021/acs.chemrev.0c01234}.

\bibitem{ref60}
Motta I, Soccio M, Guidotti G, Lotti N, Pasquinelli G.
Hydrogels for cardio and vascular tissue repair and regeneration.
\textit{Gels}. 2024;10(3):196.
\href{https://doi.org/10.3390/gels10030196}{doi:10.3390/gels10030196}.

\bibitem{ref61}
Kumar VB, Tiwari OS, Finkelstein-Zuta G, Rencus-Lazar S, Gazit E.
Design of functional RGD peptide-based biomaterials for tissue engineering.
\textit{Pharmaceutics}. 2023;15(2):345.
\href{https://doi.org/10.3390/pharmaceutics15020345}{doi:10.3390/pharmaceutics15020345}.

\bibitem{ref62}
Dashtimoghadam E, Fahimipour F, Tongas N, Tayebi L.
Microfluidic fabrication of microcarriers with sequential delivery of VEGF and BMP-2 for bone regeneration.
\textit{Sci Rep}. 2020;10(1):11764.
\href{https://doi.org/10.1038/s41598-020-68827-3}{doi:10.1038/s41598-020-68827-3}.

\bibitem{ref63}
Kong T, Zheng Q, Sun J, Wang C, Liu H, Gao Z, Qiao Z, Yang W.
Advances in magnetically controlled medical robotics: a review of actuation systems, continuum designs, and clinical prospects for minimally invasive therapies.
\textit{Micromachines}. 2025;16(5):561.
\href{https://doi.org/10.3390/mi16050561}{doi:10.3390/mi16050561}.

\bibitem{ref64}
Yang M, Zhang Y, Mou F, Cao C, Yu L, Li Z, et al.
Swarming magnetic nanorobots bio-interfaced by heparinoid-polymer brushes for in vivo safe synergistic thrombolysis.
\textit{Sci Adv}. 2023;9(48):eadk7251.
\href{https://doi.org/10.1126/sciadv.adk7251}{doi:10.1126/sciadv.adk7251}.

\bibitem{ref65}
Lin J, Cong Q, Zhang D.
Magnetic microrobots for in vivo cargo delivery: a review.
\textit{Micromachines (Basel)}. 2024;15(5):664.
\href{https://doi.org/10.3390/mi15050664}{doi:10.3390/mi15050664}.

\bibitem{ref66}
Gervasoni S, Pedrini N, Rifai T, Fischer C, Landers FC, Mattmann M, et al.
A human-scale clinically ready electromagnetic navigation system for magnetically responsive biomaterials and medical devices.
\textit{Adv Mater}. 2024;36(31):e2310701.
\href{https://doi.org/10.1002/adma.202310701}{doi:10.1002/adma.202310701}.

\bibitem{ref67}
Chen H, Zhang Y, Liu Y, Li Z, Xu J, Zhou M, et al.
Active microgel particle swarms for intrabronchial targeted delivery.
\textit{Sci Adv}. 2025;11(5):eadr3356.
\href{https://doi.org/10.1126/sciadv.adr3356}{doi:10.1126/sciadv.adr3356}.

\bibitem{ref68}
Fang Z, Yang X, Wang C, Shang L.
Microfluidics-based microcarriers for live-cell delivery.
\textit{Adv Sci (Weinh)}. 2025;12(18):e2414410.
\href{https://doi.org/10.1002/advs.202414410}{doi:10.1002/advs.202414410}.

\bibitem{ref69}
Dabravolski SA, Sukhorukov VN, Kalmykov VA, Grechko AV, Shakhpazyan NK, Orekhov AN.
The role of KLF2 in the regulation of atherosclerosis development and potential use of KLF2-targeted therapy.
\textit{Biomedicines}. 2022;10(2):254.
\href{https://doi.org/10.3390/biomedicines10020254}{doi:10.3390/biomedicines10020254}.

\bibitem{ref70}
Li YH, Kuo CH, Shi GY, Wu HL.
The role of thrombomodulin lectin-like domain in inflammation.
\textit{J Biomed Sci}. 2012;19:34.
\href{https://doi.org/10.1186/1423-0127-19-34}{doi:10.1186/1423-0127-19-34}.

\bibitem{ref71}
Nayak L, Lin Z, Jain MK.
"Go with the flow": how Krüppel-like factor 2 regulates the vasoprotective effects of shear stress.
\textit{Antioxid Redox Signal}. 2011;15(5):1449--61.
\href{https://doi.org/10.1089/ars.2010.3647}{doi:10.1089/ars.2010.3647}.

\bibitem{ref72}
Chen WH, Chen CH, Hsu MC, Chang RW, Wang CH, Lee TS.
Advances in the molecular mechanisms of statins in regulating endothelial nitric oxide bioavailability: interlocking biology between eNOS activity and L-arginine metabolism.
\textit{Biomed Pharmacother}. 2024;171:116192.
\href{https://doi.org/10.1016/j.biopha.2024.116192}{doi:10.1016/j.biopha.2024.116192}.

\bibitem{ref73}
Stepanovska B, Lange AI, Schwalm S, Pfeilschifter J, Coldewey SM, Huwiler A.
Downregulation of S1P lyase improves barrier function in human cerebral microvascular endothelial cells following an inflammatory challenge.
\textit{Int J Mol Sci}. 2020;21(4):1240.
\href{https://doi.org/10.3390/ijms21041240}{doi:10.3390/ijms21041240}.

\bibitem{ref74}
Zhang Y, Kontos CD, Annex BH, Popel AS.
Promoting vascular stability through Src inhibition and Tie2 activation: a model-based analysis.
\textit{iScience}. 2025;28(6):112625.
\href{https://doi.org/10.1016/j.isci.2025.112625}{doi:10.1016/j.isci.2025.112625}.

\bibitem{ref75}
Aoki J, Serruys PW, van Beusekom H, Ong AT, McFadden EP, Sianos G, et al.
Endothelial progenitor cell capture by stents coated with antibody against CD34: the HEALING-FIM (Healthy Endothelial Accelerated Lining Inhibits Neointimal Growth-First In Man) Registry.
\textit{J Am Coll Cardiol}. 2005;45(10):1574--9.
\href{https://doi.org/10.1016/j.jacc.2005.01.048}{doi:10.1016/j.jacc.2005.01.048}.

\bibitem{ref76}
Duckers HJ, Soullié T, den Heijer P, Rensing B, de Winter RJ, Rau M, et al.
Accelerated vascular repair following percutaneous coronary intervention by capture of endothelial progenitor cells promotes regression of neointimal growth at long term follow-up: final results of the Healing II trial using an endothelial progenitor cell capturing stent (Genous R stent).
\textit{EuroIntervention}. 2007;3(3):350--8.

\bibitem{ref77}
Choi WG, Kim SH, Yoon HS, Lee EJ, Kim DW.
Impact of an endothelial progenitor cell capturing stent on coronary microvascular function: comparison with drug-eluting stents.
\textit{Korean J Intern Med}. 2015;30(1):42--8.
\href{https://doi.org/10.3904/kjim.2015.30.1.42}{doi:10.3904/kjim.2015.30.1.42}.

\bibitem{ref78}
Ju Y, Hu Y, Yang P, Xie X, Fang B.
Extracellular vesicle-loaded hydrogels for tissue repair and regeneration.
\textit{Mater Today Bio}. 2022;18:100522.
\href{https://doi.org/10.1016/j.mtbio.2022.100522}{doi:10.1016/j.mtbio.2022.100522}.

\bibitem{ref79}
Xue J, Gurav N, Elsharkawy S, Deb S.
Hydrogel composite magnetic scaffolds: toward cell-free in situ bone tissue engineering.
\textit{ACS Appl Bio Mater}. 2024;7(1):168--81.
\href{https://doi.org/10.1021/acsabm.3c00843}{doi:10.1021/acsabm.3c00843}.

\bibitem{ref80}
Soga Y, Preul MC, Furuse M, Becker T, McDougall CG.
Calcium alginate provides a high degree of embolization in aneurysm models: a specific comparison to coil packing.
\textit{Neurosurgery}. 2004;55(6):1401--9.
\href{https://doi.org/10.1227/01.neu.0000143616.22174.67}{doi:10.1227/01.neu.0000143616.22174.67}.

\bibitem{ref81}
Maybody M, Madoff DC, Thornton RH, et al.
Catheter-directed endovascular application of thrombin: report of 3 cases and review of the literature.
\textit{Clin Imaging}. 2017;42:96--105.
\href{https://doi.org/10.1016/j.clinimag.2016.11.018}{doi:10.1016/j.clinimag.2016.11.018}.

\bibitem{ref82}
Im GB, Lin RZ.
Bioengineering for vascularization: trends and directions of photocrosslinkable gelatin methacrylate hydrogels.
\textit{Front Bioeng Biotechnol}. 2022;10:1053491.
\href{https://doi.org/10.3389/fbioe.2022.1053491}{doi:10.3389/fbioe.2022.1053491}.

\bibitem{ref83}
Blum KM, Zbinden JC, Ramachandra AB, Lindsey SE, Szafron JM, Reinhardt JW, et al.
Tissue engineered vascular grafts transform into autologous neovessels capable of native function and growth.
\textit{Commun Med (Lond)}. 2022;2:3.
\href{https://doi.org/10.1038/s43856-021-00063-7}{doi:10.1038/s43856-021-00063-7}.

\bibitem{ref84}
Schmidt CK, Medina-Sánchez M, Edmondson RJ, Schmidt OG.
Engineering microrobots for targeted cancer therapies from a medical perspective.
\textit{Nat Commun}. 2020;11(1):5618.
\href{https://doi.org/10.1038/s41467-020-19322-7}{doi:10.1038/s41467-020-19322-7}.

\bibitem{ref85}
Zhang F, Guo Z, Li Z, Luan H, Yu Y, Zhu AT, et al.
Biohybrid microrobots locally and actively deliver drug-loaded nanoparticles to inhibit the progression of lung metastasis.
\textit{Sci Adv}. 2024;10(24):eadn6157.
\href{https://doi.org/10.1126/sciadv.adn6157}{doi:10.1126/sciadv.adn6157}.

\bibitem{ref86}
Babeer A, Oh MJ, Ren Z, Liu Y, Marques F, Poly A, et al.
Microrobotics for precision biofilm diagnostics and treatment.
\textit{J Dent Res}. 2022;101(9):1009--14.
\href{https://doi.org/10.1177/00220345221087149}{doi:10.1177/00220345221087149}.

\bibitem{ref87}
Ma M, Zou F, Abudureheman B, Han F, Xu G, Xie Y, et al.
Magnetic microcarriers with accurate localization and proliferation of mesenchymal stem cell for cartilage defects repairing.
\textit{ACS Nano}. 2023;17(7):6373--86.
\href{https://doi.org/10.1021/acsnano.2c11320}{doi:10.1021/acsnano.2c11320}.

\end{thebibliography}
\end{document}